\documentclass[]{aa}
\usepackage{graphicx}
\usepackage{txfonts}

\begin{document}

\title{Spectral line width decrease in the solar corona: resonant energy conversion from Alfv{\'e}n to acoustic waves}

\author{T.V. Zaqarashvili\inst{1,2}, R. Oliver\inst{1} \and J.L. Ballester\inst{1}}

\institute{Departament de F\'{\i}sica,
Universitat de les Illes Balears, E-07122 Palma de Mallorca, Spain, \\
\email{[temury.zaqarashvili;ramon.oliver;dfsjlb0]@uib.es} \and
Georgian National Astrophysical Observatory (Abastumani
Astrophysical
Observatory), Kazbegi Ave. 2a, Tbilisi 0160, Georgia, \\
\email{temury@genao.org}}

\offprints{T. Zaqarashvili}

\date{Received / Accepted }

\abstract{Observations reveal an increase with height of the line width of several coronal
spectral lines probably caused by outwardly propagating Alfv{\'e}n waves.
However, the spectral line width sometimes shows a sudden decrease at a height
$\sim$ 0.1--0.2 $R_{\sun}$, where the ratio of sound to Alfv{\'e}n speeds may approach unity.}{To explain the
observed line width reduction in terms of the energy conversion from Alfv{\'e}n to another type of wave
motion.}{Weakly non-linear wave-wave interaction in ideal MHD.}{Qualitative analysis shows that the resonant energy conversion from Alfv{\'e}n to acoustic waves near the region of the corona where the plasma $\beta$ approaches unity may explain the observed spectral line width reduction.}{}


\titlerunning{Spectral line width decrease by resonant Alfv{\'e}n wave energy conversion}
\authorrunning{Zaqarashvili et al.}

\maketitle

\section{Introduction}

The non-thermal broadening of coronal spectral lines was first observed by a
rocket-borne instrument (Hassler et al. \cite{has}) and more recently by the CDS and SUMER
instruments on the SOHO spacecraft (Banerjee et al. \cite{ban}; Doyle et al. \cite{doy1,doy2}; Harrison et al.
\cite{har}; O'Shea et al. \cite{osh1,osh2}). It was
found that the Doppler width generally increases with height and this was
interpreted as a signature of outwardly propagating undamped Alfv{\'e}n waves.
However, some observations (Harrison et al. \cite{har}; O'Shea et al. \cite{osh1,osh2}) show a sudden decrease of the
line width at an approximate height of 0.1--0.2 $R_{\sun}$ above the solar
surface. These observations have been performed in coronal structures with different properties
(e.g. on the poles and the equator).

It is very important to understand the mechanism of line width
decrease as it may trigger the acceleration of plasma particles in
these regions. In polar coronal holes, where the magnetic field is
open and predominantly vertical, Alfv{\'e}n waves mainly contribute
to the off-limb line broadening due to their transverse velocity
polarisation. Acoustic waves propagating along the magnetic field
are unlikely to contribute to the line broadening because their
velocity polarisation is predominantly perpendicular to the line of
sight. Then the decrease of the line width in polar coronal holes
can be explained either by the Alfv{\'e}n wave damping or due to the
conversion into acoustic waves. It was first suggested by Hollweg
(\cite{hol}) that the ponderomotive force of linearly or
elliptically polarised Alfv{\'e}n waves may lead to the resonant
generation of acoustic waves when the Alfv{\'e}n and sound speeds
become similar (or $\beta=8\pi p/B^2 \sim 1$, where $p$ is the
plasma pressure and $B$ is the magnetic field). Recently,
Zaqarashvili and Roberts (\cite{zaq1}) presented a detailed scenario
of wave coupling for $\beta \sim 1$ and showed that Alfv{\'e}n and
sound waves alternately exchange energy when they propagate along a
uniform magnetic field with the same phase speed. This means that
the Alfv{\'e}n wave resonantly transfers energy into acoustic waves
in a $\beta \sim 1$ plasma and {\it vice versa}.

However, the value of the plasma $\beta$ varies with height in the solar atmosphere: it is greater than unity at the
photospheric level, but it quickly
decreases upwards and becomes smaller in the corona. Nevertheless, it increases again
with height and in the solar wind generally $\beta \geq 1$. Therefore, there
must be a region in the corona where $\beta$ approaches unity. Recent modelling
of the plasma $\beta$ above an active region (Gary \cite{gary}) shows that this
parameter approaches unity at relatively low coronal heights, of the order of
$0.2 R_{\sun}$ (see Fig. 3 in Gary's paper), which surprisingly coincides with the
height at which the line width decrease is observed.

Thus, Alfv{\'e}n waves propagating upwards from the coronal base and entering the
region $\beta \sim 1$ can resonantly transfer energy into sound waves, which consequently leads to the observed decrease of the spectral line width near the same height.

In this letter, we present a qualitative analysis of the wave conversion process and compare to the observed decrease of the line width.

\section{Energy conversion from Alfv{\'e}n to acoustic waves}

We study this process in an open magnetic field configuration
using the non-linear, ideal magnetohydrodynamic (MHD) equations written in terms
of the fluid velocity, ${\bf u}$, the magnetic field, $\bf B$, the pressure,
$p$, and the density, $\rho$. Gravity is neglected in the equations, as it does
not significantly affect the wave interaction process in the corona.

We use a Cartesian coordinate system with the $z$-axis directed radially outward
from the Sun. The unperturbed magnetic field is directed along the $z$-axis,
i.e. ${\bf B}_0=(0,0,B_0)$, and the unperturbed pressure and density, $p_0$ and
$\rho_0$, are allowed to vary in the $z$-direction.

Let us consider wave propagation strictly along the magnetic field, i.e. along
the $z$-axis. Then the parallel and perpendicular components of the MHD
equations take the form

\begin{eqnarray}
{{{\partial {\rho}}}\over {\partial t}} + {\rho}{{{\partial
{u_{\parallel}}}}\over {\partial z}} + u_{\parallel}{{{\partial {\rho}}}\over {\partial
z}}=0, \label{pert_eq1} \\
{\rho}{{{\partial u_{\parallel}}}\over {\partial t}} + {\rho}u_{\parallel}{{{\partial
{u_{\parallel}}}}\over {\partial z}}=-{{{\partial p}}\over {\partial z}}-{{\partial}\over {\partial z}}{{b^2_{\perp}}\over {8\pi}},\\
{\rho}{{{\partial u_{\perp}}}\over {\partial t}} + {\rho}u_{\parallel}{{{\partial
{u_{\perp}}}}\over {\partial z}}={{B_0}\over {4\pi}}{{\partial b_{\perp}}\over
{\partial z}},\\
{{{\partial b_{\perp}}}\over {\partial t}}= B_0{{{\partial u_{\perp}}}\over {\partial z}}-u_{\parallel}{{{\partial b_{\perp}}}\over
{\partial z}}-b_{\perp}{{{\partial u_{\parallel}}}\over {\partial z}},\\
{{{\partial p}}\over {\partial t}} + {\gamma}p{{{\partial
{u_{\parallel}}}}\over {\partial z}} + u_{\parallel}{{{\partial p}}\over {\partial
z}}=0, \label{pert_eq5}
\end{eqnarray}

\noindent where $p$ and $\rho$ denote the total pressure and density,
$u_{\parallel}$ and $u_{\perp}$ are the parallel and perpendicular components of
the velocity, while $b_{\perp}$ is the magnetic field perturbation.  These
equations describe the fully non-linear behavior of adiabatic sound and
Alfv{\'e}n waves propagating along an applied magnetic field. The total plasma
density and pressure are the sums of the unperturbed and  perturbed parts,
$\rho_0 + \rho_1$ and $p_0 + p_1$, respectively. Then $u_{\parallel}$, $\rho_1$
and $p_1$ are the sound wave velocity, density, and pressure perturbations,
whereas $u_{\perp}$ and $b_{\perp}$ are the Alfv{\'e}n wave velocity and
perturbed magnetic field component. We next express perturbations as

\begin{eqnarray}
\rho_1(z,t)={\tilde \rho}_1(z)e^{i\omega_\mathrm{s} t} +
{\tilde \rho}^{*}_1(z)e^{-i\omega_\mathrm{s} t}, \label{pert1} \\
u_{\parallel}(z,t)={\tilde u_{\parallel}}(z)e^{i\omega_\mathrm{s} t} +
{\tilde u_{\parallel}}^{*}(z)e^{-i\omega_\mathrm{s} t},\\
p_1(z,t)={\tilde p}_1(z)e^{i\omega_\mathrm{s} t} + {\tilde
p}^{*}_1(z)e^{-i\omega_\mathrm{s} t},\\
b_{\perp}(z,t)={\tilde b_{\perp}}(z)e^{i\omega_\mathrm{A} t} +
{\tilde b_{\perp}}^{*}(z)e^{-i\omega_\mathrm{A} t},\\
u_{\perp}(z,t)={\tilde u_{\perp}}(z)e^{i\omega_\mathrm{A} t} +
{\tilde u_{\perp}}^{*}(z)e^{-i\omega_\mathrm{A} t}, \label{pert4}
\end{eqnarray}

\noindent where $\omega_\mathrm{s}$ and $\omega_\mathrm{A}$ are the sound and Alfv{\'e}n wave
frequencies, correspondingly, and the symbol $*$ denotes the complex conjugate.

Substitution of expressions (\ref{pert1})--(\ref{pert4}) into equations
(\ref{pert_eq1})--(\ref{pert_eq5}) and subsequent time averaging over a time
interval much larger than the sound and Alfv\'en wave periods leads to the
cancellation of second order terms, so that only the linear and third order
advective terms remain. This means that Alfv{\'e}n and sound waves are decoupled
and that the energy exchange process does not occur. Nevertheless, if the wave
frequencies satisfy the condition $\omega_\mathrm{s}=2\omega_\mathrm{A}$, then the second order
terms also remain and we obtain a set of non-linear differential equations with
$z$-derivatives describing the non-linear energy exchange between sound and
Alfv{\'e}n waves. Let us consider the weakly non-linear process in which
perturbations are much smaller than the unperturbed values. Then, using the
adiabatic relation $p_1=c^2_\mathrm{s}{\rho}_1$ and neglecting third and higher order
terms leads to the equations

\begin{eqnarray}
{{\partial {\tilde \rho}_1}\over {\partial z}}= - {{2i\omega_\mathrm{A}
\rho_0}\over {c^2_\mathrm{s}}} {\tilde u_{\parallel}}- {{{2c^{\prime}_\mathrm{s}}\over
{c_\mathrm{s}}}}{\tilde \rho}_1-{{i\omega_\mathrm{A} \rho_0}\over
{B_0c^2_\mathrm{s}}}{\tilde b_{\perp}} {\tilde u_{\perp}}, \label{pert_eq1b} \\
{{\partial {\tilde u_{\parallel}}}\over {\partial z}}=
- {{2i\omega_\mathrm{A}}\over {\rho_0}}{\tilde \rho}_1
- {{{{\rho}^{\prime}_0}\over {\rho_0}}}{\tilde u_{\parallel}}, \label{pert_eq2b} \\
{{\partial {\tilde u_{\perp}}}\over {\partial z}}= {{i\omega_\mathrm{A}}\over
{B_0}}{\tilde b_{\perp}}-{{2i\omega_\mathrm{A}}\over {\rho_0B_0}}{\tilde
\rho}_1{\tilde b_{\perp}}^{*} - {{i\omega_\mathrm{A}}\over {v^2_\mathrm{A}}}{\tilde
u_{\parallel}}{\tilde u_{\perp}}^{*}-
{{{{\rho}^{\prime}_0}\over {\rho_0B_0}}}
{\tilde u_{\parallel}}{\tilde b_{\perp}}^{*}, \label{pert_eq3b} \\
{{\partial {\tilde b_{\perp}}}\over {\partial z}}= {{i\omega_\mathrm{A}
B_0}\over {v^2_\mathrm{A}}}{\tilde u_{\perp}}-{{i\omega_\mathrm{A} B_0}\over
{\rho_0v^2_\mathrm{A}}}{\tilde \rho}_1{\tilde u_{\perp}}^{*}-
{{i\omega_\mathrm{A}}\over {v^2_\mathrm{A}}}{\tilde u_{\parallel}}{\tilde
b_{\perp}}^{*}, \label{pert_eq4b}
\end{eqnarray}

\noindent where $c_\mathrm{s}$ and $v_\mathrm{A}$ are the sound and Alfv{\'e}n speeds and primes
denote the derivative with respect to $z$. Equations
(\ref{pert_eq1b})--(\ref{pert_eq4b}) describe the time-averaged, weakly
non-linear spatial behavior of Alfv{\'e}n and sound waves propagating along an
applied magnetic field.

\begin{figure}
\centering
\includegraphics[width=1\linewidth]{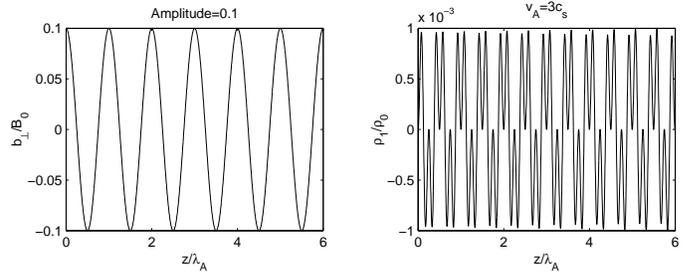}
\caption{Spatial dependence of the magnetic field and density perturbations for
$v_\mathrm{A}=3c_\mathrm{s}$. At $z=0$ there are only Alfv{\'e}n waves with an amplitude ${\tilde
b_{\perp}}/B_0=0.1$. The amplitude of Alfv{\'e}n waves remains unchanged with height: there is no energy
transfer into acoustic waves.}
\end{figure}

Let us, for simplicity, first consider an isothermal, homogeneous medium with
uniform $\rho_0$ and $p_0$.  If the backreaction of sound waves is neglected
(which means that the non-linear terms in equations
(\ref{pert_eq3b})--(\ref{pert_eq4b}) are neglected), then we get the equation
for linear Alfv{\'e}n waves, which leads to the solutions ${\tilde u_{\perp}},\,
{\tilde b_{\perp}} \sim e^{i k_\mathrm{A} z}$ and the dispersion relation
$\omega^2_\mathrm{A}=v^2_\mathrm{A}k^2_\mathrm{A}$, where $k_\mathrm{A}$ is the wave number of Alfv{\'e}n waves.
Thus, in this case the amplitude of Alfv{\'e}n waves remains constant over the
whole spatial domain (this is because we neglected the backreaction of sound
waves). Then, the sound wave equations (\ref{pert_eq1b}) and (\ref{pert_eq2b})
give

\begin{equation}
{{\partial^2 {\tilde u_{\parallel}}}\over {\partial z^2}} +
{{4\omega^2_\mathrm{A}}\over {c^2_\mathrm{s}}}{\tilde u_{\parallel}}= -
{{2\omega^3_\mathrm{A}}\over {B^2_0k_\mathrm{A}c^2_\mathrm{s}}}{\tilde b^2_{\perp}}. \label{pert_eq_sound}
\end{equation}

\noindent This equation is typical of oscillations under the action of an
external force and is characterised by having a resonant solution when the
wavelength of the external oscillations coincides with the natural oscillatory
wavelength of the system. Here the ponderomotive force of Alfv{\'e}n waves acts
as an external periodic driver. Therefore, the sound wave resonant harmonics
possess a wave number $2k_\mathrm{A}$ (because of the square of $b_{\perp}$ in equation
[\ref{pert_eq_sound}]) when $v_\mathrm{A} \approx c_\mathrm{s}$ is satisfied. Therefore, the
ponderomotive force of Alfv{\'e}n waves may resonantly drive sound waves in the
vicinity of the $\beta \sim 1$ region, an effect which to our knowledge was
first pointed out by Hollweg (\cite{hol}).

\begin{figure}
\centering
\includegraphics[width=1\linewidth]{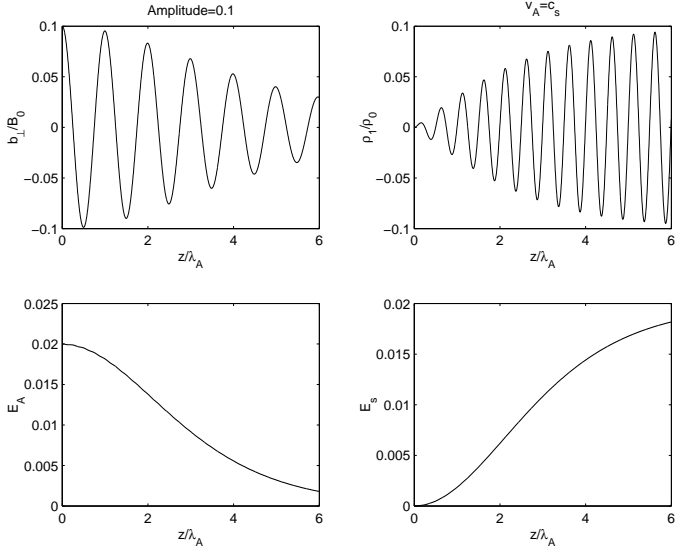}
\caption{ Spatial dependence of the magnetic field and density
perturbations and energy of Alfv{\'e}n and acoustic waves under
resonant conditions, i.e. for $v_\mathrm{A}=c_\mathrm{s}$. At $z=0$
there are only Alfv{\'e}n waves with an amplitude ${\tilde
b_{\perp}}/B_0=0.1$. Because of the resonant process, the energy of
Alfv{\'e}n waves decays with height and density perturbations with
half the wavelength of the Alfv{\'e}n wave are exponentially
amplified.}
\end{figure}

However, for a comparison with observations, the backreaction of
sound waves must be taken into account in order to see the decrease
of Alfv{\'e}n wave amplitude due to the energy conversion. We
therefore solve numerically the complete set of equations
(\ref{pert_eq1b})--(\ref{pert_eq4b}) with boundary conditions
representing the upward propagation of Alfv\'en waves from the
coronal base. We thus impose $\tilde b_\perp/B_0=-\tilde
u_\perp/v_\mathrm{A}\neq 0$, $\tilde u_\parallel=\tilde\rho_1=0$ at
$z=0$.

The spatial dependence of the magnetic field and density perturbations for
$v_\mathrm{A}=3c_\mathrm{s}$ is shown in Fig. 1. At $z=0$ there are only Alfv{\'e}n waves with a
relative amplitude ${\tilde b_{\perp}}/B_0=0.1$ and a wavelength $\lambda_\mathrm{A}$,
while all perturbations related to acoustic waves are zero. For $z>0$ the
ponderomotive force excites density fluctuations with a  wavelength smaller than
$\lambda_\mathrm{A}$, but the amplitude of Alfv{\'e}n waves is unchanged. In addition,
density fluctuations remain small and display no spatial growth, while the
modulation with half the wavelength of Alfv{\'e}n waves can be traced. Thus,
there is no resonant energy transfer from Alfv{\'e}n into acoustic waves. This
is a typical behavior of a forced system when the external frequency of
oscillation does not coincide with the system's natural frequency of oscillation
(in the present context the word ``frequency'' should be substituted by
``wavelength'').

\begin{figure}
\centering
\includegraphics[width=1\linewidth]{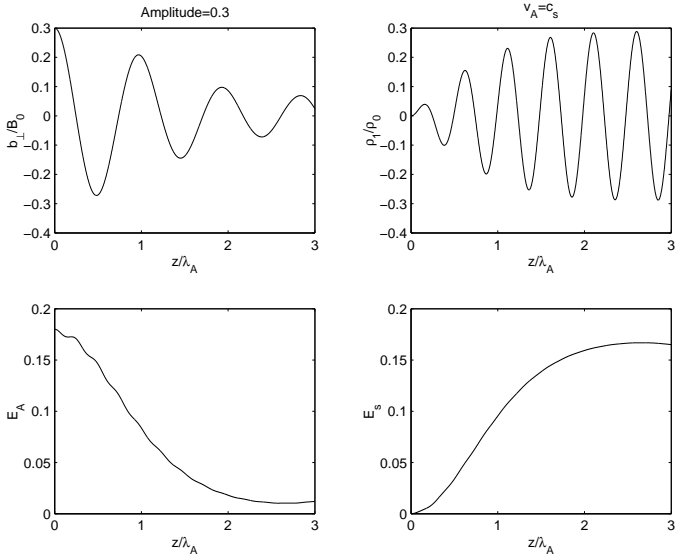}
\caption{Same as in Fig. 2, but for the Alfv{\'e}n wave amplitude
${\tilde b_{\perp}}/B_0=0.3$ at $z=0$. The rapid decay of the
Alfv{\'e}n wave in 2--3 wavelengths is clearly seen.}
\end{figure}

Next let us consider the case $v_\mathrm{A}=c_\mathrm{s}$. The spatial dependence of the magnetic
field and density perturbations and the energies of Alfv{\'e}n and acoustic
waves are plotted in Fig. 2. At low heights (i.e. for small $z$) there are only
Alfv{\'e}n waves with an amplitude ${\tilde b_{\perp}}/B_0=0.1$ and acoustic
wave perturbations are negligible. At larger heights the energy of Alfv{\'e}n
waves decreases and density perturbations with half the wavelength of Alfv{\'e}n
waves are exponentially amplified. It is clearly seen that the energy of
Alfv{\'e}n waves is effectively transformed into sound wave energy. Moreover,
the efficiency of the energy conversion depends on the wave amplitude and the
stronger the amplitude of Alfv{\'e}n waves at $z=0$, the faster the energy
conversion process. To show this effect, in Fig. 3 we plot the results for
$v_\mathrm{A}=c_\mathrm{s}$, but now with a larger Alfv{\'e}n wave amplitude at $z=0$, namely
${\tilde b_{\perp}}/B_0=0.3$. The faster decay of the Alfv{\'e}n wave amplitude
is evident: the amplitude of Alfv{\'e}n waves decreases by almost a factor
three in a distance of two wavelengths.

In the case of a spatially inhomogeneous plasma $\beta$ it can be
shown that Alfv{\'e}n waves propagating with almost constant
amplitude in the $\beta \not=1$ region undergo a rapid decay
(transferring their energy into density perturbations) when they
enter the region $\beta \sim 1$. In this case the height variation
of unperturbed parameters (density, temperature, magnetic field)
relevant to real conditions in the corona must be given. However
this is not the scope of this letter and is thus left for a future
study.

Following Gary (\cite{gary}), the plasma $\beta$ above an active
region can approach unity at relatively low coronal heights ($z\sim
0.2 R_{\sun}$). Outside active regions $\beta$ can approach unity
even at lower heights. Hence, Alfv{\'e}n waves propagating outward
from the solar coronal base can efficiently decay and their energy
transformed into sound waves in or near the regions where the plasma
$\beta$ is close to unity. In polar coronal holes, where a sudden
decrease of the Doppler width has been observed (O'Shea et al.
\cite{osh1,osh2}), the open magnetic field is almost vertical.
Therefore only the velocity component transverse to the field lines
may significantly contribute to the off-limb line broadening. On the
contrary, the velocity component along the field lines will cause a
negligible contribution because it is almost perpendicular to the
line of sight. Hence the conversion of Alfv{\'e}n into acoustic
waves can be the reason for the observed decrease of the line width
in coronal holes. This suggestion is further supported by the
agreement between the heights where the spectral line width decrease
is observed and where $\beta$ approaches unity. In the ideal case,
the energy transferred to acoustic waves will be returned back to
Alfv\'en waves after some distance, although this may not be the
case in a more realistic situation since the generated sound waves
can be quickly attenuated by Landau damping (or other damping
mechanism) leading to the acceleration of plasma particles.

However if the magnetic field is inclined or has a complex structure
(for example, above active regions) then the excited acoustic waves
may have a velocity component parallel to the line of sight. Then
they also will contribute to the line broadening and therefore the
line width will not be reduced. This is further supported by the
fact that  some observations do not show the line width reduction
(Wilhelm et al. \cite{wil}). So, possibly, the observation of line
width reduction depends on the magnetic field orientation in the
observed region.

\subsection{Comparison with observations}

Now let us make a comparative analysis between the theoretically obtained decay
due to energy conversion and the observed decrease of the line width.
Observations (Harrison et al. \cite{har}; O'Shea et al. \cite{osh1,osh2}) show a similar decay rate of the line width:
this quantity decreases by $\sim 10\%$ over a distance of $\sim$ 70--100 Mm.
This means that the amplitude of Alfv{\'e}n waves is also reduced by $\sim 10\%$
over the same distance. In order to estimate the decay rate of Alfv{\'e}n waves,
their wavelength must be known. Unfortunately, observations do not reveal
neither the wavelength nor the period of propagating Alfv{\'e}n waves, although
some estimations can be done. Let us suppose that the Alfv{\'e}n wave period is
$\sim$ 5 min. In the region where $\beta \sim 1$, the Alfv{\'e}n and sound
speeds have similar values, i.e. $\sim$ 200 km/s (thus, in this region $v_\mathrm{A}$ is
smaller than its value in the lower corona, $\sim 1000$ km/s). Then, from
$v_\mathrm{A}\sim 200$ km/s and assuming a wave period $\sim$ 5 min, the wavelength is
$\lambda_\mathrm{A}\sim 60$ Mm. So if the line broadening is caused by the damping of
Alfv{\'e}n waves with a period of 5 min, then their amplitude is reduced by
$\sim 10\%$ over a distance of 1--2$\lambda_\mathrm{A}\sim$ 60--120 Mm, in good agreement
with observations. The observed spectral line width implies Alfv{\'e}n wave
velocity perturbations of the order of 20--40 km/s (Harrison et al. \cite{har}). Thus, the
relative amplitude (wave velocity divided by phase speed) of Alfv{\'e}n waves
can be estimated as 0.1--0.2. Hence, for a comparison let us take the
theoretical plot for Alfv{\'e}n waves with relative amplitude of 0.1, i.e. Fig.
2. This figure shows that the Alfv{\'e}n wave amplitude is reduced by $10\%$ in
1--1.5 wavelengths, which is in perfect agreement with observations.

\section{Conclusions}

We show that the resonant energy conversion from Alfv{\'e}n to sound
waves near the region where the plasma $\beta$ approaches unity (or
more precisely, where the ratio of sound to  Alfv{\'e}n speeds
approaches unity) may explain the observed sudden decrease of the
spectral line width in the solar corona. The estimated decay rate of
Alfv{\'e}n waves with period $\sim$ 5 min and amplitude 0.1
perfectly fits the observations. However, Alfv{\'e}n waves with
shorter period and smaller amplitude (or longer period and larger
amplitude, see Fig. 3) can also explain the observed decay with the
same success. Hence, the determination of the period or wavelength
of propagating Alfv{\'e}n waves responsible for the reported line
broadening will be a good test for this theory.

Sound waves (or more properly, ion-acoustic waves) may accelerate
plasma particles due to Landau damping, which is very efficient when
the electron temperature approaches the ion temperature, because the
wave phase speed now resides well inside the Maxwellian distribution
of particle velocities (Chen \cite{chen}), therefore the generated
ion-acoustic waves will be quickly damped, transforming their energy
into kinetic energy of plasma particles. As a result, the solar wind
acceleration may begin near the region of the solar corona where the
hydrodynamic and magnetic energy densities approach each other. This
requires detailed observational and theoretical study.

\section{Acknowledgements}

The work was supported by MCyT grant AYA2003-00123.

\end{document}